\documentclass[manuscript,screen]{acmart}
\usepackage{multirow}
\usepackage{bbm}
\usepackage{amsmath,bm}
\usepackage[normalem]{ulem}
\usepackage{paralist}

\definecolor{blue-violet}{rgb}{0.54, 0.17, 0.89}

\usepackage{subcaption}
\usepackage{tikz}
\usepackage{makecell}
\usepackage{enumitem}
\usetikzlibrary{matrix}

\usepackage[mode=buildnew]{standalone}

\copyrightyear{2024}
\acmYear{2024}
\setcopyright{rightsretained}
\acmConference[FAccT '24]{The 2024 ACM Conference on Fairness, Accountability, and Transparency}{June 3--6, 2024}{Rio de Janeiro, Brazil}
\acmBooktitle{The 2024 ACM Conference on Fairness, Accountability, and Transparency (FAccT '24), June 3--6, 2024, Rio de Janeiro, Brazil}\acmDOI{10.1145/3630106.3658968}
\acmISBN{979-8-4007-0450-5/24/06}
\begin{document}

\title{The Dark Side of Dataset Scaling: Evaluating Racial Classification in Multimodal Models}

\author{Abeba Birhane}
  \authornote{Equal contribution}
\email{birhanea@tcd.ie}
\orcid{0000-0001-6319-7937}
\affiliation{%
  \institution{Mozilla Foundation \& School of Computer Science and Statistics, Trinity College Dublin}
  \city{Dublin}
  \country{Ireland}
}
\author{Sepehr Dehdashtian}
  \authornote{Equal contribution}
\email{sepehr@msu.edu}
\orcid{0000-0002-2512-8815}
\affiliation{%
  \institution{Michigan State University, Department of Computer Science and Engineering}
  \city{Michigan}
  \country{USA}
}
\author{Vinay Uday Prabhu}
\email{vinaypra@alumni.cmu.edu}
\orcid{0000-0001-9602-0134}
\affiliation{%
  \institution{HAL51 Inc}
  \city{San Francisco, California}
  \country{USA}
}
\author{Vishnu Boddeti}
\email{vishnu@msu.edu}
\orcid{0000-0002-8918-9385}
\affiliation{%
  \institution{Michigan State University, Department of Computer Science and Engineering}
  \city{Michigan}
  \country{USA}
}

\begin{abstract}
`Scale the model, scale the data, scale the GPU farms' is the reigning sentiment in the world of generative AI today. While model scaling has been extensively studied, data scaling and its downstream impacts on model performance remain under-explored. This is particularly important in the context of multimodal datasets whose main source is the World Wide Web, condensed and packaged as the Common Crawl dump, which is known to exhibit numerous drawbacks. In this paper, we evaluate the downstream impact of dataset scaling on  14 visio-linguistic models (VLMs) trained on the LAION400-M and LAION-2B datasets by measuring racial and gender bias using the Chicago Face Dataset (CFD) as the probe. Our results show that as the training data increased, the probability of a pre-trained CLIP model misclassifying human images as offensive non-human classes such as chimpanzee, gorilla, and orangutan decreased, but misclassifying the same images as human offensive classes such as criminal increased. Furthermore, of the 14 Vision Transformer-based VLMs we evaluated, the probability of predicting an image of a Black man and a Latino man as \textit{criminal} increases by \textit{65\%} and \textit{69\%}, respectively, when the dataset is scaled from 400M to 2B samples for the larger ViT-L models. Conversely, for the \textit{smaller} base ViT-B models, the probability of predicting an image of a Black man and a Latino man as \textit{criminal} decreases by \textit{20\%} and \textit{47\%}, respectively, when the dataset is scaled from 400M to 2B samples.  We ground the model audit results in a qualitative and historical analysis, reflect on our findings and their implications for dataset curation practice, and close with a summary of mitigation mechanisms and ways forward. All the meta-datasets curated in this endeavor and the code used are shared at: \url{https://github.com/SepehrDehdashtian/the-dark-side-of-dataset-scaling}.\\
{\color{red}\textit{Content warning: This article contains racially dehumanising and offensive descriptions.}}
\end{abstract}


\maketitle

\section{Introduction}
\label{sec:intro}

Over the past few years, transformer-based models have come to revolutionize the field of deep learning. These models leverage a mechanism known as attention or self-attention~\cite{vaswani2017attention}, allowing them to weigh the influence of different input elements and capture long-range dependencies. Transformer models have been instrumental in tasks such as text and speech translation~\cite{kano2021transformer,le2020dual}, genomic research~\cite{gen0_zhang2022transformer,gen1_lee2022learning,gen2_kang2020learning,gen3_choi2023transformer}, anomaly detection in time series~\cite{an0_tuli2022tranad,an1_chen2021learning}, and fraud detection~\cite{fraud0_wang2023financial,fraud1_chen2021improving,fraud2_yang2023finchain}. Furthermore, the principles of transformer models have been extended to other domains, such as computer vision, with the introduction of Vision Transformers (ViT). The parallel processing capability of these models has significantly improved the efficiency of both training and inference~\cite{karita2019comparative}. The landscape of transformer models is dynamic, with continuous advancements contributing to the state-of-the-art where notable models include RoBERTa~\cite{liu2019roberta}, XLNet~\cite{xlnet_yang2019xlnet}, and GPT-4~\cite{achiam2023gpt}.

However, the 
ubiquitous development and adoption of artificially intelligent (AI) technologies -- including transformer models -- into numerous societal domains has also 
ushered in a multitude of actual and potential risks and harms. Some of the most notable crises in current AI include functional failures~\cite{raji2022fallacy}; disparate performance and treatment based on gender, race, and other dimensions~\cite{obermeyer2019dissecting,chu2022digital,imana2021auditing}; 
exacerbation of discriminatory, stereotypical and otherwise marginalising values~\cite{luccioni2023stable,barlas2021see,scheuerman2019computers}; 
legal incompatibility, plagiarism, and copyright violations~\cite{kilovaty2019legally,patronusai24}; privacy violations~\cite{mireshghallah2023can,cadwalladr2018revealed,veliz2021privacy}; production and spread of misinformation~\cite{kim2024fables,chen2023can}; massive consumption of energy and resources~\cite{crawford2021atlas,luccioni2023power};
opaque and inscrutable datasets, models and practices~\cite{pasquale2015black,metaxa2021auditing}; power centralization~\cite{birhane2022values,abdalla2021grey}; the normalization of surveillance~\cite{browne2015dark,kalluri2020don}; and labour exploitation~\cite{gray2019ghost,tubaro2019micro}.  Continual and comprehensive documentation, checks, critical scrutiny, evaluations, and testing of these systems have become pressing. Subsequently, audits of algorithmic systems -- including of datasets -- have emerged as one of the most effective mechanisms for diagnosing, documenting, and mitigating numerous AI risks and harms. 

Mislabeling and misclassification of people's images, particularly of those from minoritized groups has been one of the major problems in computer vision systems.  
In 2015, Google's Photo app classified photos of Jacky Alciné and his friend (both of whom are Black) as ``gorillas''~\cite{kasperkevic2015google,simonite2018comes}. Eight years later in 2023, the problem remains unsolved~\cite{Grant2023}. There has since been growing awareness of racial and gender bias in computer vision and multimodal models, which has ushered in significant improvements in the accuracy of image classification.  
Yet, misclassification of images, particularly of minoritized races and genders remains one of the most persistent problems. Over the past years, a robust body of work has demonstrated the tendency of machine learning systems, 
tools, and applications to encode and exacerbate societal stereotypes and historical biases~\cite{benjamin2019race,noble2018algorithms,browne2015dark,mcquillan2022resisting,mandal2023multimodal,birhane2021large}.

In 2018,~\citet{buolamwini2018gender} evaluated three commercial classification algorithms along the dimensions of gender and skin tone. In what has now become one of the canonical studies that paved the way for algorithmic auditing as a field of study, they found 
statistically significant disparities in performance showing up to 34.7\% error rate for dark-skinned females, compared to an error rate of under 0.8\% for lighter-skinned males. Numerous subsequent studies have demonstrated that computer vision models often fail to detect and/or accurately classify images of genders, races, and demographics outside the status quo. For example, computer vision models failed to detect images from non-Western demographic groups~\cite{de2019does}, image detection -- in the context of pedestrian detection -- showed lower rate of pedestrian detection on darker-skin tones while exhibiting a higher rate of precision for lighter-skin tones~\cite{wilson2019predictive}, and unsupervised image representation models encode implicit racial, gender, and intersectional bias~\cite{steed2021image}.

Racial and gender bias in language models~\cite{caliskan2017semantics,abid2021persistent,bailey2022based} and vision models~\cite{buolamwini2018gender,scheuerman2019computers,keyes2018misgendering,hamidi2018gender} is a well documented phenomena. Bias in multimodal models (models with any combination of text, image, audio, and video modalities as inputs/outputs), on the other hand, are sparsely studied. Still, a rapidly growing body of work indicates that multimodal models also encode and exacerbate societal and historical stereotypes and biases, in some cases at a much worse scale than that of models with a single modality. For example~\citet{mannering2023analysing}, found gender bias in text-to-image models using object detection, ~\citet{luccioni2023stable} found that outputs from diffusion models encode societal biases,~\citet{hendricks2018women} found that race and gender bias are exacerbated in downstream outputs in image captioning, \citet{mandal2023multimodal} found that DALL-E 2 and Stable Diffusion reflect gender bias, and~\citet{bianchi2022easily} found that image generation models encode and exacerbate societal and historical stereotypes along with complex biases in generated images. More specifically, \citet{hundt2022robots} audited 
the multimodal model CLIPort~\cite{shridhar2021cliport}, which runs on a robot and is backed by CLIP, for 
performance on terms such as 
``criminal", ``homemaker", and ``doctor" on the eight variants of race and gender in the CFD and found significant bias and negative stereotypes. 
Similarly, ~\citet{liu2024scoft} trained text to image generators, diffusion models in particular, on a small dataset of images and descriptions collected by residents of different countries, then generated images on multiple approaches. Humans rated the models on offensiveness, stereotypes, image to description match, and cultural representativeness. 
They found that CLIP cosine similarity scores get worse as the models improve on each of the aforementioned human rated metrics.

The drive towards Artificial General Intelligence (AGI) being pursued by various actors entails a 
triadic interplay between the verticals of computing power, model architecture and data. Thus far, constructing a defensible \textit{moat} with computing power and model architecture advancements alone has proven elusive. Most of the key players in Big Tech and in the startup ecosystem alike source their computing resources from the same set of two or three key players that supply the silicon and orchestrate the cloud computing software. On the model architecture front, the marquee architectures have rarely been disrupted, with almost all the major players training on some variant of the U-Net-like, ResNet-like or transformer-like architectures. It is in the data vertical that players try to establish their idiosyncratic moats, often 
scraping and munging data from the unsuspecting corners of the internet that are not guarded by authorization checks~\cite{WebScrap67:online}, pirated book-dumps~\cite{Microsof77:online} and even fandom wikis, casino wholesaling websites, and even random internet comments~\cite{nasr2023scalable} which is the considered their secret sauce and guarded fiercely. 
In the specific context of visio-linguistic models (VLMs), 
one canonical 
source for 
datasets  
has been the Common Crawl (CC) repository,  
a collection of periodically web crawled data maintained by a San Francisco based 501(c)(3) non–profit organization. This primary source has been distilled to generated datasets such as~\href{https://laion.ai/blog/large-openclip/}{LAION-400M and LAION-5B}. 
The recipe entails using a pre-trained black-box VLM (typically a variant of the CLIP~\cite{clip_radford2021learning} model published by OpenAI) purportedly to filter images whose alt-text description closely resemble their semantic content. For example, the \textit{plain-vanilla} CLIP model and its ViT B/32 CLIP variant were used to distill the CC dataset into LAION-400M with $0.3$ and $0.28$ cosine similarity thresholds, respectively. In this study, we investigate the questions of what happens to the quality of such distilled datasets whose scale is increased by expanding coverage of the CC data-dump and manipulating the ad hoc hand-set cosine similarity thresholds as well as the subsequent downstream effects of the models' predictions trained on these datasets.

We audit pre-trained Contrastive Language-Image Pretraining (CLIP)~\cite{clip_radford2021learning} models for gender and ethnicity bias. Specifically, we evaluate 14 Vision Transformer-based VLMs from OpenCLIP~\cite{ilharco_gabriel_2021_5143773_openclip} on a classification task using the Chicago Face Dataset (CFD)~\cite{ma2015chicago} as a probe dataset. CLIP model architectures comprise two encoders: a vision transformer for image inputs and a transformer-based language model for text inputs. These encoders project the input data into a shared embedding space, enabling the model to compare and relate visual and textual information. One of the key features of CLIP is its ability to perform zero-shot learning. This means that the model can ``generalize" to tasks it has not been explicitly trained on, such as prediction of a novel class. This is achieved by leveraging the flexibility of natural language as a prediction space. However, CLIP models suffer numerous problems. The CLIP paper~\cite{clip_radford2021learning} itself (in Section 7.1) outlined that images belonging to the ``Black" racial designation had an approximately 14\% chance of being miscategorized as \texttt{[‘animal’, ‘gorilla’, ‘chimpanzee’, ‘orangutan’, ‘thief’, ‘criminal’ and ‘suspicious person’]} in their \textit{FairFace} dataset experiment.  
We replicate the Zero-Shot CLIP experiment using the CFD (see Appendix~\ref{fig:cfd-images} for a sample of hand blurred images) as a probe dataset and study the effect of scaling the pre-training dataset. Our analysis finds that the effect of scaling the datasets is dependent on the scale of the trained model. Larger models exhibit a greater proclivity towards predicting specific racial groups like Black and Latin faces as criminals as the scale of the pre-training datasets increases. On the other hand, smaller models exhibit a lower proclivity towards predicting specific racial groups like Black and Latin faces as criminals as pre-training datasets' scale increases.

\section{Audit Methodology}
\label{sec:method}

To quantitatively evaluate the downstream consequences of scaling the pre-training datasets, we first 
explored model variants where the architecture was held constant and two or more model checkpoints were provided: some trained with LAION-400M and some trained with LAION-2B-en. The emergence of OpenCLIP~\cite{ilharco_gabriel_2021_5143773_openclip} facilitated this endeavor as (to the best of our knowledge) it remains the only resource that hosts VLM variants with \textit{fixed model architecture} but varying dataset sizes (LAION-400M and LAION-2B-en respectively). Among the 120 models present in OpenCLIP (at the time of our experimentation), we selected the following 14 CLIP-model pairs presented in Table~\ref{tab:openclip_variants} that met our criteria. The architecture of the models that we evaluated, their pre-training dataset, the number of parameters of each architecture, and the number of \textbf{FL}oating point \textbf{OP}eration\textbf{s} (FLOPs) are listed in Table~\ref{tab:openclip_variants}. The OpenCLIP project currently uses an idiosyncratic naming convention for the model checkpoints presented in the second column of Table~\ref{tab:openclip_variants}. To evaluate the effect of scaling the pre-training dataset on these model variants, we used the 
CFD~\cite{ma2015chicago}, as a probe. We replicated the \textit{Zero-Shot CLIP experiment} that appeared in \textit{Section 7.1-Bias} of the original CLIP paper~\cite{clip_radford2021learning} by OpenAI, the details of which are in Subsection~\ref{sec:cfd_exp_details}. 
\begin{table}[ht!]
\caption{Architecture-Dataset variants in the OpenCLIP ecosystem we evaluated in this study. 
\label{tab:openclip_variants}}
\centering
\begin{tabular}{llrr}
\toprule
Architecture & Dataset/Checkpoint       & \makecell{Number of\\Parameters (M)} & FLOPs (B) \\
\midrule
\multirow{3}{*}{ViT-B-16}      
    &   \texttt{laion400m\_e31}        &  \multirow{3}{*}{149.62}  &  \multirow{3}{*}{41.09} \\
    &   \texttt{laion400m\_e32}        &    &   \\
    &   \texttt{laion2b\_s34b\_b88k}   &    &   \\
\midrule
\multirow{2}{*}{ViT-B-16-plus-240}      
    &   \texttt{laion400m\_e31}   &  \multirow{2}{*}{208.38}  &  \multirow{2}{*}{64.03} \\
    &   \texttt{laion400m\_e32}   &    &   \\
\midrule
\multirow{4}{*}{ViT-B-32}      
    &   \texttt{laion400m\_e31}        &  \multirow{4}{*}{151.28}  &  \multirow{4}{*}{14.78} \\
    &   \texttt{laion400m\_e32}        &    &   \\
    &   \texttt{laion2b\_e16}          &    &   \\
    &   \texttt{laion2b\_s34b\_b79k}   &    &   \\
\midrule
\multirow{2}{*}{ViT-B-32-quickgelu}      
    &  \texttt{laion400m\_e31}         &  \multirow{2}{*}{151.28}  &  \multirow{2}{*}{14.78} \\
    &  \texttt{laion400m\_e32}         &    &   \\
\midrule
\multirow{3}{*}{ViT-L-14}      
    &   \texttt{laion400m\_e31}                           &  \multirow{3}{*}{427.62}  &  \multirow{3}{*}{175.33} \\
    &   \texttt{laion400m\_e32}        &    &   \\
    &   \texttt{laion2b\_s32b\_b82k}        &    &   \\
\bottomrule
\end{tabular}
\end{table}                                                    

The CFD is a highly controlled dataset that consists of high resolution images of 597 unique individuals along with their \textit{self-classified} race and gender labels belonging to Asian (109), Black (197), Latin (108), and White (183) categories.  
A (blurred) sample of images and additional information on the CFD dataset is shown in Appendix~\ref{fig:cfd-images}. The dataset has been meticulously standardized to control 
for potentially confounding causal covariates such as facial expressions, resolution, image-pixel saturation, lighting conditions, clothing, and eye gaze. The 597 images have each individual wearing the same heather grey t-shirt. While much smaller in volume, unlike the  
the majority of openly available datasets, the individuals in CFD had their consent obtained, were financially compensated 
and were given the option to self-classify from a set of pre-defined options: from Black, White, Asian or Latin and Female or Male.

\subsection{Experiment Design}
\label{sec:cfd_exp_details}

The sub-phases involved in the bias analysis experiments 
were as follows:
\\\textbf{1: Image pre-processing}: All the 597 images with neutral expressions extracted from CFD were pre-processed using the respective OpenCLIP model's built-in \texttt{preprocess} function that entails resizing (to size $224\times224$), center-cropping and pixel intensity normalization sub-processes. The output of this sub-phase is a CFD-image-tensor, $\mathbf{I}_{cfd} \in \mathbb{R}^{597\times 224 \times 224 \times 3}$.%
\\\textbf{2: Class-generation and tokenization}: We first created an 8-class vector with the following classes \texttt{[‘human being’, ‘animal’, ‘gorilla’, ‘chimpanzee’, ‘orangutan’, ‘thief’, ‘criminal’, and ‘suspicious person’]}. Except for the \texttt{`human being'} class, which was added by us, the remaining seven classes were verbatim extracted from \textit{Section 7.1 Bias} of the OpenAI CLIP paper~\cite{clip_radford2021learning}. Next, we created the class-sentences using the \texttt{"A photo of a/an <class>}" template\footnote{As advocated in the \texttt{Interacting with CLIP} Jupyter notebook shared at \url{https://github.com/mlfoundations/open_clip/blob/main/docs/Interacting_with_open_clip.ipynb} in the context of \textit{Zero-Shot Image Classification for CIFAR-100 dataset}. These eight sentences were then \textit{tokenized} using OpenCLIP's tokenizer module (the \texttt{Vocab size} is 49408 for all the models considered in this paper), thus yielding an $8 \times 77$ sized token-matrix.}. The output of this sub-phase is a sparse zero-padded text-token matrix, $\mathbf{T}_{8-class} \in \mathcal{I}^{ 8\times 77}$ where $\mathcal{I}=[0,..., N_{tokens}-1]$ is the tokenizer-index set. 
\\\textbf{3: Forward pass, feature extraction and normalization}: The pre-processed image tensors and the text-tokens generated in the previous sub-phase were now fed into the encoder of the chosen OpenCLIP model, and the output image and text features were then normalized.  For
all the evaluated 
models, the features are 768-dimensional thus rendering the text and image feature matrices over the 597 neutral-expression CFD images to be $597 \times 768$. That is, the image-feature matrix is $\mathbf{F}_{I}=\left[\mathbf{f}_{0}^{I},...,\mathbf{f}_{596}^{I}\right]^T\in \mathbb{R}^{597\times 768} $ 
and the text-feature matrix would is $\mathbf{F}_{\tau}=\left[\mathbf{f}_{0}^{t},...,\mathbf{f}_{7}^{t}\right]^T\in \mathbb{R}^{768\times 8}$.

To highlight how self-similar the $8 \times 8$ textual features are, we present the annotated heatmap of the $\mathbf{F}_\tau \times \mathbf{F}_\tau^T$ matrix (see Figure~\ref{fig:self_sim}(a) in the Appendix). Similarly, we also present the heatmap of the $597 \times 597$ sized $\mathbf{F}_I \times \mathbf{F}_I^T$ matrix (Appendix Figure~\ref{fig:self_sim}(b)). Given the fact that the 597 images were sorted and grouped by Race-Gender categories, the block-like structures visible (in the Appendix Figure~\ref{fig:self_sim}(b)) indicate the fact that the model's output features are certainly 
influenced by these categorical indicators. 
\\\textbf{4: Computing softmax-matrices}: Firstly, we obtain the image-text cosine similarity matrix, $\mathbf{C} \in \mathbb{R}^{597 \times 8}$ as:
\begin{equation}
\mathbf{C}=\mathbf{F}_{I}\mathbf{F}_{\tau}^T.
\label{eq:cosine_mat}
\end{equation}
Then, the softmax-matrix $\mathbf{S} \in \mathcal{P}^{597 \times 8}$ ($\mathcal{P}=\left\{p | 0<p<1 \right\}$) is computed as: 
\begin{equation}
\mathbf{S}=\text{softmax}\left(100\times \mathbf{C} \right).
\label{eq:softmax_mat}
\end{equation}
Here $softmax()$ is the softmax function applied row-wise. That is, if $\mathbf{C}_{i,j}$ is the $i^{th}$ row $j^{th}$ column element in the cosine-matrix, then the corresponding $(i,j)^{th}$ element in the softmax-matrix, $\mathbf{S}_{i,j}$ would be ${\boldsymbol S}_{\mathbf i\boldsymbol,\mathbf j}=\;\frac{\exp\left(100\times{\mathbf C}_{i,j}\right)}{\sum_{k=0}^7\exp\left(100\times{\mathbf C}_{i,k}\right)}$.

\section{Results\label{sec:cfd_results}}

\begin{figure}
    \centering
    \includegraphics[width=\linewidth]{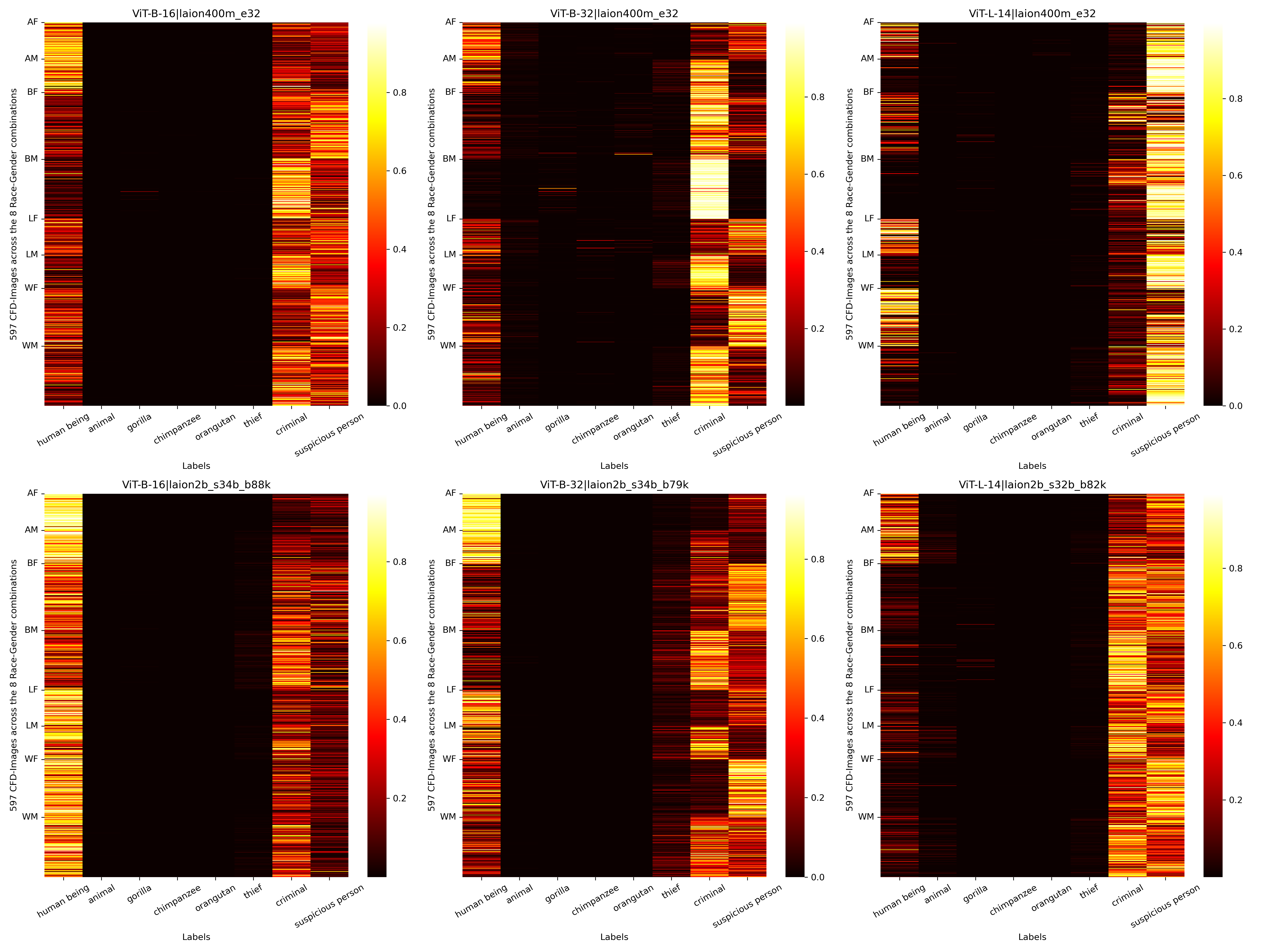}
    \vspace{-0.3cm}
    \caption{Heatmaps of 597 × 8 softmax-matrices for three models (columns) and two pre-training datasets (rows).\label{fig:results:heatmap-images}}
\end{figure}

In Figure~\ref{fig:results:heatmap-images} we first present the three 597 × 8 sized output softmax-matrices obtained from ViT-B-16, ViT-B-32, and ViT-L-14 with two different pre-training datasets: 
LAION-400M and LAION-2B-en.  
The $(i, j)^{th}$ element of each of these matrices captures the softmax score value of the $j^{th}$ class $(j \in \{0, ..., 7\})$ obtained from that specific OpenCLIP model in response to the $i^{th}$ input CFD image $(i \in \{0, ..., 596\})$. The 597 rows (representing the 597 CFD images) are grouped by their self-classified Race-Gender groupings. That is, the first 57 rows represent images from the Asian-Female (abbreviated as AF), and the next 52 rows map to the Asian Male (AM) group, and so on. The titles of these subplots are formatted as strings with two fields separated by the ‘|’ character: <model architecture> | <pre-training dataset/checkpoint>. From the figure, we make the following observations. 

\begin{figure}
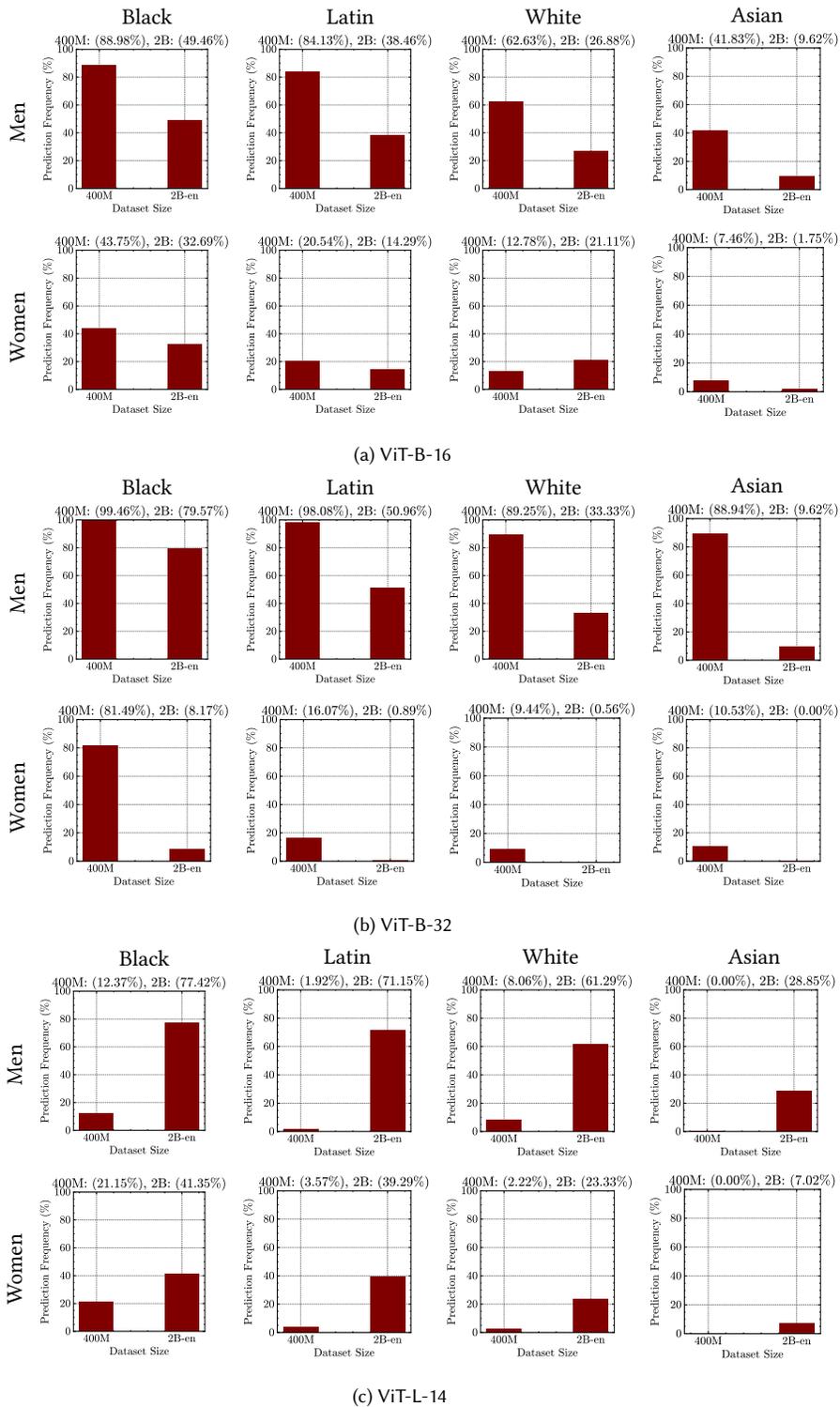

    \centering
    \begin{subfigure}[c]{0.75\linewidth}
        \centering
        \input{figs/barplots/barplot-B-16}
        \caption{ViT-B-16\label{fig:results:barplots-B-16}}   
    \end{subfigure}
    \begin{subfigure}[c]{0.75\linewidth}
        \centering
        \input{figs/barplots/barplot-B-32}
        \caption{ViT-B-32\label{fig:results:barplots-B-32}}   
    \end{subfigure}
    \begin{subfigure}[c]{0.75\linewidth}
        \centering
        \input{figs/barplots/barplot-L-14}
        \caption{ViT-L-14\label{fig:results:barplots-L-14}}   
    \end{subfigure}
    \vspace{-0.3cm}
    \caption{Effect of scaling the dataset from 400M to 2B on the frequency of an image from CFD getting predicted as `criminal' for each race-gender group and three different architectures: ViT-B-16 (a), ViT-B-32 (b), and ViT-L-14 (c). We observe that the larger ViT-L model's predilection for labeling faces as `criminals' increases significantly for black and Latino men when the pre-training dataset is scaled from 400M to 2B (see Section~\ref{sec:cfd_results}, specifically 3.1 to 3.3). }
    \label{figs:results:barplots}
\end{figure}

\begin{figure}[tb]
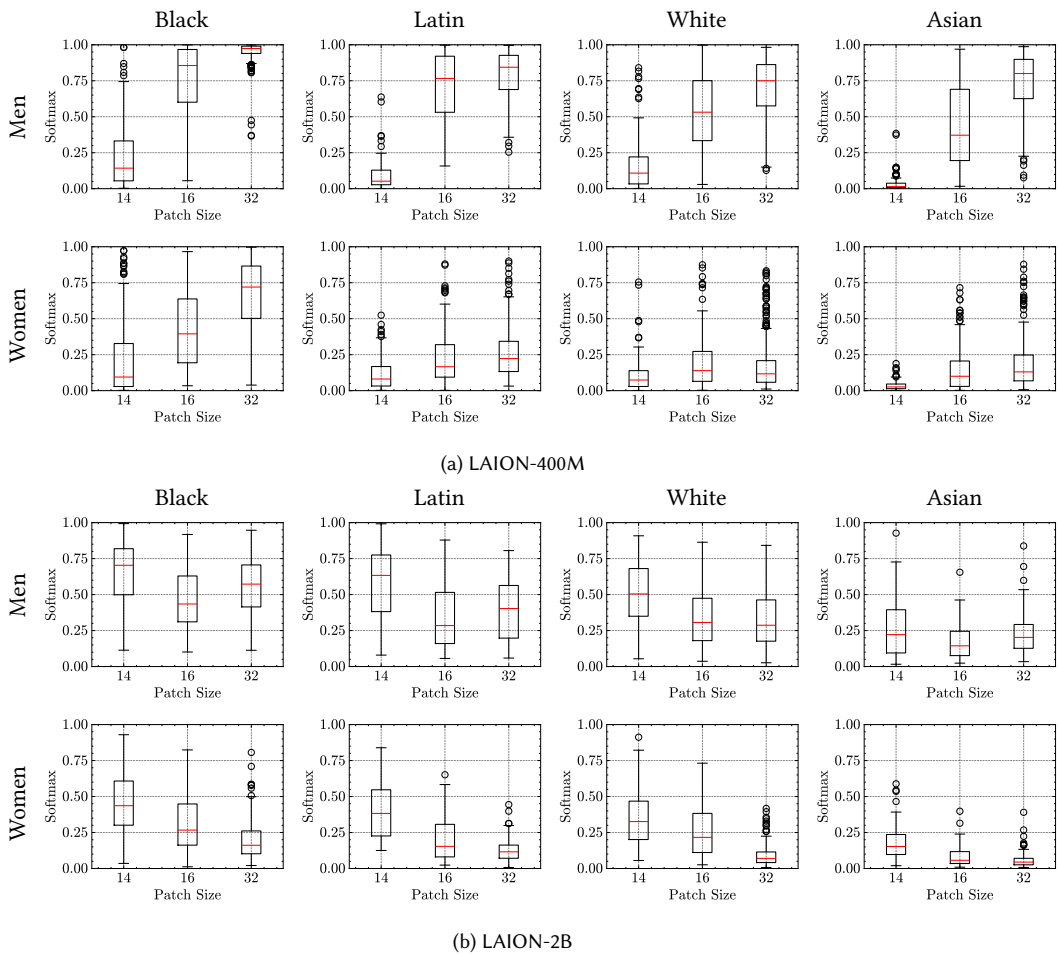

    \centering
    \begin{subfigure}[c]{0.9\linewidth}
        \centering
        \input{figs/vsPatch/boxplot/boxplot-400M}
        \caption{LAION-400M\label{fig:results:boxvsPatch-400M}}   
    \end{subfigure}
    \begin{subfigure}[c]{0.9\linewidth}
        \centering        \input{figs/vsPatch/boxplot/boxplot-2B}
        \caption{LAION-2B\label{fig:results:boxvsPatch-2B}}   
    \end{subfigure}
    \caption{Plots showing the effect of \textbf{patch size} on the distribution of ``criminal'' predictions for (a) LAION-400M and (b) LAION-2B as the pre-training datasets.}
    \label{figs:results:boxvsPatch}
    \vspace{-0.2cm}
\end{figure}

\begin{figure}
    \centering
    \captionsetup[subfigure]{oneside,margin={0em,3.7em}}
    \begin{subfigure}[c]{0.49\linewidth}
        \centering\includegraphics[width=0.9\linewidth]{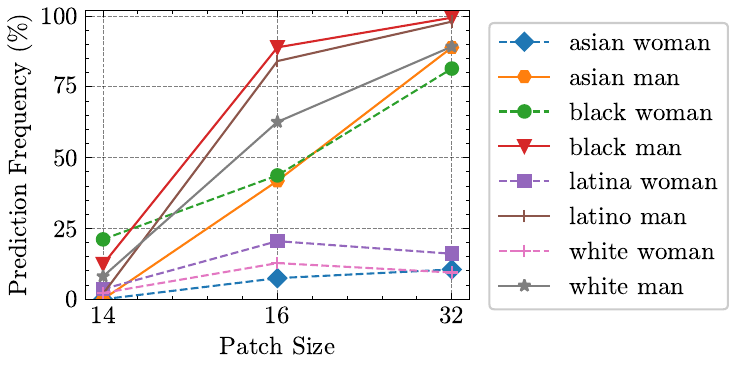}
        \caption{LAION-400M\label{fig:results:meanvsPatch-400M}}   
    \end{subfigure}
    \begin{subfigure}[c]{0.49\linewidth}
        \centering\includegraphics[width=0.9\linewidth]{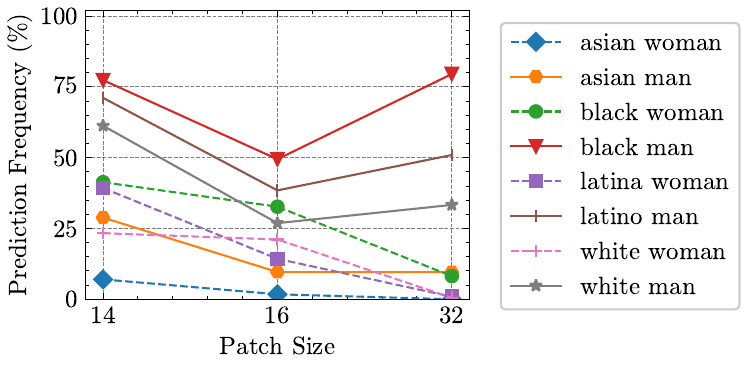}
        \caption{LAION-2B\label{fig:results:meanvsPatch-2B}}   
    \end{subfigure}
    \vspace{-0.3cm}
    \caption{Frequency of `criminal' prediction versus patch size for the LAION-400M (a) and for the LAION-2B (b) datasets.}
    \label{figs:results:meanvsPatch}
\end{figure}

\noindent\textbf{1. Non-human offensive labels:} For all the models we evaluated, regardless of the training data size (LAION-400M or LAION-2B), the softmax scores for non-human offensive labels i.e., $animal$, $gorilla$, $chimpanzee$, and $orangutan$ are close to zero across different architectures and datasets. In other words, none of the models accurately predicted images of people 
from CFD with the `human being' class. Instead the models predicted these images of humans with the other non-human offensive classes: $animal$, $gorilla$, $chimpanzee$, and $orangutan$.\\
\noindent\textbf{2. Human being label:} We found that, as training data size increased from 400M to 2B samples, model accuracy at predicting human faces from CFD accurately as \textit{human being} increased for all races and genders and by \textbf{6.4\%} for Black women and \textbf{58\%} for Asian men.\\ 
\noindent\textbf{3. Human offensive labels:} Among the three human offensive labels ($thief$, $criminal$, and \emph{suspicious person}), we found \emph{criminal} and \emph{suspicious person} predictions occur the most. The scores for the \emph{suspicious person} class increase as the number of model parameters is scaled from ViT-B-16 (149.62 M) to ViT-L-14 (427.62 M). We also observe that the scores for the \emph{human being} class increase for ViT-B-16 and ViT-B-32 when the training dataset is scaled to 2B samples, while they decrease for ViT-L-14 when the same scaling is applied. For women (of all four races) the probability of an image being predicted as $P_{\text{thief}}$ is zero and for men (of all 4 races), this probability is almost zero, with an average prediction of 0.01. Generally, we found that men are more likely to be classified as \emph{criminal} than women (see Figure~\ref{figs:results:barplots}). Furthermore, for the \emph{criminal} class, we found:
\begin{enumerate}[label={\bfseries 3.\arabic*}]\item \textbf{Scaling increases $\emph{criminal}$ prediction:} As shown in Figure~\ref{figs:results:barplots}, for the smaller models, i.e. (a) ViT-B-16 and (b) ViT-B-32, scaling the number of samples from 400M to 2B in the pre-training datasets decreased the \emph{criminal} prediction by \textbf{33\% }on average. However, for the large models (c) (ViT-L-14), increasing the amount of pre-training data from 400M to 2B samples, increased the probability of an image of a person from CFD being predicted as \emph{criminal} by \textbf{37.5\% }on average. 
For all four races of the CFD human images, for large models (ViT-L-14 in Figure~\ref{figs:results:barplots} (c)), the \emph{criminal} label was allocated to \textbf{Latin (71\%)} and \textbf{Black (77\%)} faces at a higher rate compared to the White (61\%) and Asian (28\%) groups. For all models we evaluated, \textbf{Black} and \textbf{Latin} groups received higher probabilities of being predicted as \emph{criminal} compared to the other two groups: White and Asian. For example, for the ViT-B-16 model pre-trained on the LAION-400M dataset, \emph{criminal} is \textbf{66\%} and \textbf{52\% }for \textbf{Black} and \textbf{Latin} faces respectively, while it is \textbf{37\%} and \textbf{24\%} for \textbf{White} and \textbf{Asian} faces, respectively. This was the case regardless of dataset scale, meaning that the probability of the label \emph{criminal} being allocated to Black and Latin racial groups was \textbf{highest} for models trained on both 400M and 2B samples. 

\item\textbf{Patch size increased \emph{criminal} prediction:} We found that, within the same racial group, men are generally labeled as \emph{criminal} at a higher rate (\textbf{45\% higher} on average) than women. As shown in Figure~\ref{figs:results:boxvsPatch} (a), for models trained on the smaller dataset, increased model patch size increases the probability of an image of a face being predicted as \emph{criminal} for all racial and gender groups \textbf{except for White women}. However, for models trained on the larger 2B sample, the probability of \emph{criminal} decreased for women, as the patch size of the model increased. 

\item\textbf{Frequency of \emph{criminal} prediction versus patch size:} As shown in Figure~\ref{figs:results:meanvsPatch}, the frequency of an image from CFD being labeled as \emph{criminal} increased for all groups except White and Latina women for all models trained on 400M samples (Figure~\ref{figs:results:meanvsPatch} (a)). In other words, the models showed bias against Black and Asian women, and all men from the four racial groups (Black, Asian, Latin and White). In other words, we see a close to 100\% prediction frequency for the model with patch size 32 for Black men, which means that the \textit{model predicts all images of Black men} from the CFD as \emph{criminal}. Conversely, the frequency of \emph{criminal} prediction decreased for women (of all racial groups) for models trained on the 2B sample (Figure~\ref{figs:results:meanvsPatch} (b)), as the patch size of the model increased. This means that fewer women were classified as \emph{criminal} when the patch size of the model increased. 
\end{enumerate}

\begin{figure}
    \centering\includegraphics[width=\linewidth]{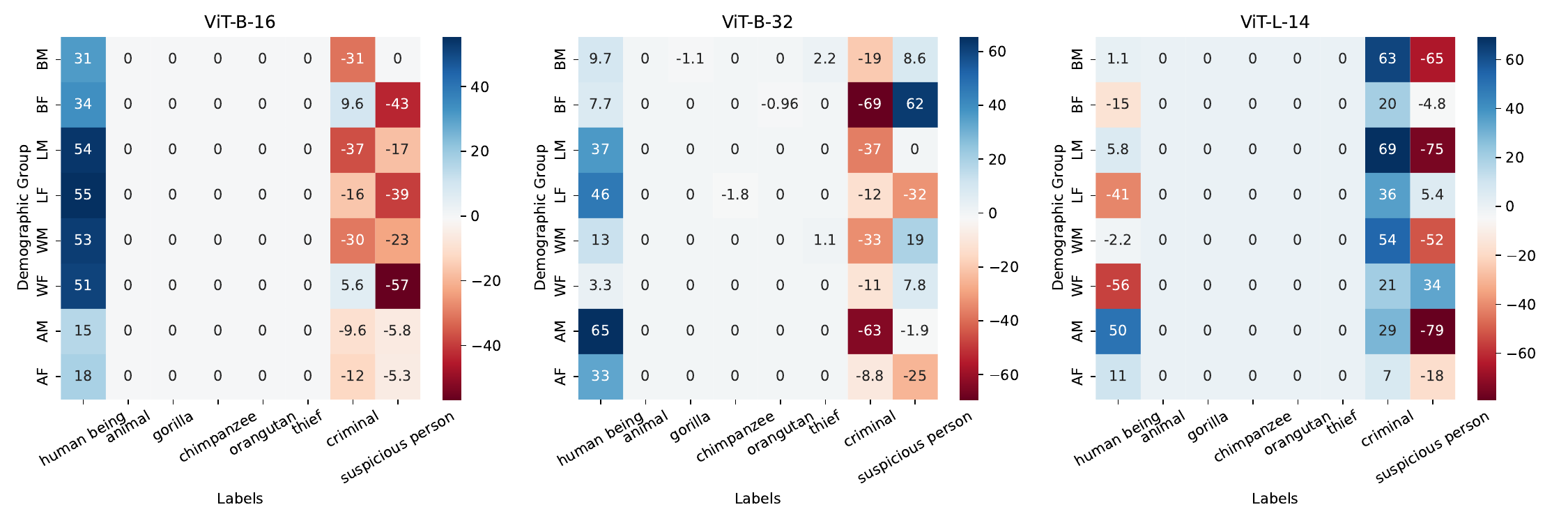}
    \vspace{-0.5cm}
    \caption{The effect of dataset scaling on the predictions of the models. The numbers show the change in probabilities when the pre-trained dataset is scaled from 400M to 2B for ViT-B-16, ViT-B-32, and ViT-l-14. Positive values mean an increase in the probability when the number of pre-training samples is increased. All values are in percentage. }
    \label{figs:results:heatmap-summary}
\vspace{-0.3cm}
\end{figure}

\noindent\textbf{4. Summary of the effect of dataset scaling on models' predictions:} We summarize the effect on the models' predictions as we scale the pre-training dataset, shown in Figure~\ref{figs:results:heatmap-summary}. For all racial groups, the probability of an image of a human from the CFD being predicted as \emph{human being} was higher in smaller models (ViT-B-16 and ViT-B-32). On the other hand, the probability of an image of a human from the CFD being predicted as \emph{human being} decreased for Latino women, Black women, White women, and White men in the larger models (ViT-L-14). For the larger ViT-L-14 models, the heatmaps demonstrate the disparate increase in the probability of labeling human faces as \emph{criminal} across different racial groups. Similarly, for the smaller ViT-B models, the heatmaps also demonstrate the disparate decrease in the probability of labeling faces as \emph{criminal} across different racial groups. We also found that, in general, Latino/Latina individuals were misclassified with high confidence as one of the `Asian' classes and this\textit{ misclassification increased with dataset scaling} (see Appendix~\ref{fig:crosstab_rg} for details). All of our results, as well as the meta-dataset created as a result of our audit, are available on our~\href{https://github.com/SepehrDehdashtian/the-dark-side-of-dataset-scaling}{repository}. 

\vspace{-0.5cm}

\section{Qualitative Analysis: Dehumanization and Criminalization of Black Bodies\label{sec:dehuman}}

A rich body of work within Science and Technology Studies (STS), Black studies, and critical data and algorithm studies has emphasized the tendency of ML research, tools, and applications to encode and exacerbate societal stereotypes and historical injustice~\cite{benjamin2019race,noble2018algorithms,browne2015dark,mcquillan2022resisting}. As presented in Section~\ref{sec:cfd_results}, our findings extend this rich body of work by demonstrating that not only do large VLMs encode such 
historical trend that dehumanizes Black bodies but also, as these models and datasets increase in scale, such dehumanization is further exacerbated.

During the institution of slavery, Black people, particularly Black men, were depicted as 
``brute'' and ``docile'' creating and reinforcing the idea that the most fitting position for them is slavery~\cite{smiley2016brute}. Scientific racism enabled  racial classification 
in 18th and 19th century that justified slavery, legal segregation and discrimination~\cite{saini2019superior}. Arbitrary racial classifications emerged that portrayed Caucasians (white Europeans) as the epitome of humanity at the top of the hierarchy, while these arbitrary systems placed African and African-Americans at the bottom of the racial hierarchy. Although practices such as chattel slavery and legal segregation were eventually abolished, systemic racism -- which is rooted in these vacuous underlying conceptions -- remain embedded in institutions, social structures and processes and continue to be pervasive and ingrained in societal systems~\cite{feagin2013systemic,elias2024brief}. 

In the U.S., the rise of the for-profit prison industrial complex is a primary example that embodies systemic racism maintaining the cycle of systemic racism through unjust incarceration of Black people and unrealistic depictions of Black men as ``thug'', ``criminal'', and ``suspicious''
~\cite{smiley2016brute,feagin2013systemic}. 
Many prison companies mandate that municipalities have a 90-95\% prison occupancy rate increasing targeted association of Black people and crime~\cite{alexander2020new,bardes2018redefining}. Such stereotypes and racist ideologies have fueled racial violence, criminalization, and mass incarnation of Black men, especially in the US. Black bodies, according to~\cite{bey2016bring}, are often perceived as a threat and typecast as ``gangster,'' ``rapist'', and ``ghetto''. The ``Black-as-criminal'' stereotype, subsequently, can result in non-violent acts of Black men being perceived as violent and aggressive while violent acts performed by white men are perceived as unintentional or get attributed to external factors and uncontrollable causes such as mental health~\cite{chapple2017blacklivesmatter}. 

Contrary to these racial stereotypes, a robust body of work, especially in the context of the U.S., documents that Black men commit crimes at a far lower rate than whites, while Black people constitute the group that are victims of violent crimes at far higher rates than whites~\cite{gross2022race,gaston2019enforcing}. Innocent Black people, according to~\citet{gross2022race}, are seven-and-a-half times more likely to be convicted of murder than whites, and convicted Black people are 80\% more likely to be innocent than other convicted murderers. 
In 2002, Black people were six times more likely to be murdered than whites, and this number was much higher during previous decades, where 47\% of victims were African Americans during the 1976-2002 period~\cite{rosich2007race}. Conversely, a 2018 United Nations Report on racial disparities ~\cite{sentencing2018report} shows that ``African-American adults are 5.9 times as likely to be incarcerated than whites'' and more likely than whites to be arrested; once arrested, more likely to be convicted; and once convicted, more likely to be incarnated than whites. Studies on drug use across demographers in the US reveal a similar trend. 
Although African Americans and whites use illegal drugs at similar rates, Black people are 19 times more likely to be convicted of drug crimes than whites~\cite{gross2022race,rosich2007race}. 

\begin{figure}[tb]
    \centering
    \includegraphics[width=\columnwidth]{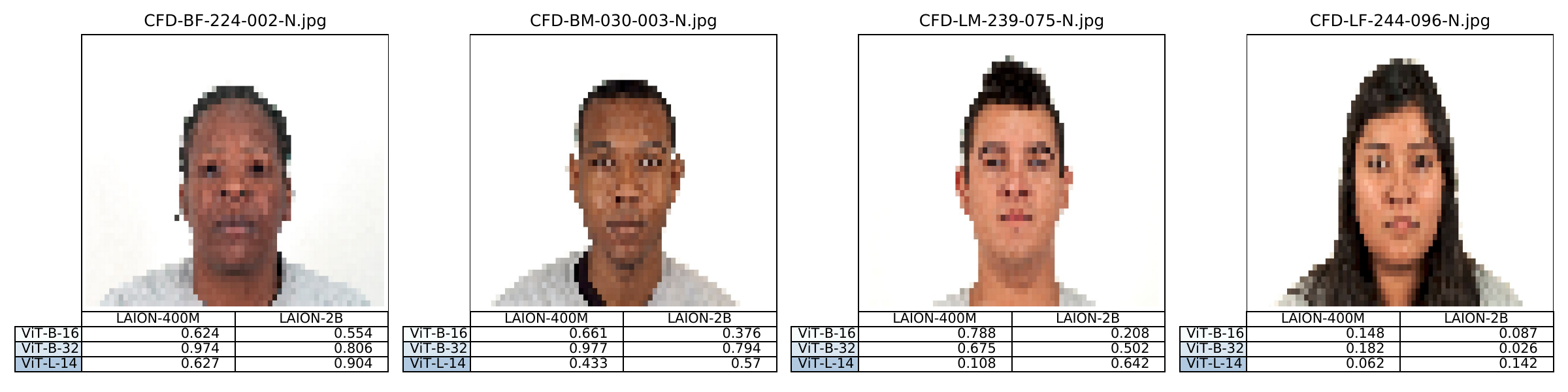}
    \caption{Example images of Black individuals from CFD  and the tendency of the OpenCLIP models studied to associate them with the ``A photo of a criminal” sentence. The title(s) indicates the file name, and the table under each image shows $P_{\text{criminal}}$ for three different architectures and two pre-training datasets. The images are hand-blurred to preserve the privacy of the data subjects.}
    \label{fig:CFD_criminal}
\vspace{-0.3cm}
\end{figure}

Erroneous stereotypes have historically (and currently) served to explicitly, implicitly, and systematically place Black people, particularly Black men, as ``suspects'', ``criminals'', or ``persons of interest''~\cite{smiley2016brute}. Along with past work that has highlighted the risk of models amplifying racial stereotypes~\cite{bianchi2022easily,van2016stereotyping,scheuerman2020we}, our findings confirm this trend. As outlined in Section~\ref{sec:cfd_results}, we observe that current SoTa models encode and exacerbate racial stereotypes. Furthermore, as outlined in~\ref{figs:results:meanvsPatch}, the likelihood of a Black man 
being classified as ``criminal'' 
\textit{increases as training datasets get bigger}. As illustrated in Figure~\ref{figs:results:meanvsPatch} (a), the prediction frequency for Black men as ``criminal'' for model patch size 32 was close to 100\%, where all the CFD samples for Black men were predicted as ``criminal''. (See Figure~\ref{fig:CFD_criminal}, for a randomly selected example images of four Black individuals from the CFD dataset each showing criminality prediction for three different model architectures). As reported in the tables, the association of Black and Latin faces with \texttt{`A photo of a criminal'} increases for the large model (ViT-L-14) while it decreases for smaller models such as ViT-B-16 and ViT-B-32 by scaling the dataset from 400M to 2B. To summarise, the findings from our audits align with the rich body of work within Black studies, critical data studies, and critical race scholarship that have examined, underscored and challenged systemic racism. To that end, as training datasets get larger, they further exacerbate deeply ingrained negative societal and historical stereotypes and racial dehumanisation, particularly against Black people.     

\section{Discussion and Recommendations}
\label{sec:recommendations}
In this paper, we have systematically examined the impact of dataset scaling and model architecture by evaluating 14 Vision Transformer-based VLMs pre-trained on two datasets: LAION 400M and LAION-2B-en. Our results show that with a larger ViT-L model the predilection for labeling faces as `criminals' increased statistically significantly for black and Latino men when the pre-training dataset was scaled from 400M to 2B samples. Datasets are fundamental backbones to models and party determine whether a model is equitable, just, robust, and well-performing. Subsequently, a transparent and just sourcing and rigorous evaluation, audit, curation, and management of datasets is critical for advancing the field towards a healthy and sustainable direction. 

We  
strongly highlight the need to avoid interpreting the empirical results from a reductionist lens where the emphasis is erroneously laid on the specific details of the metrics introduced (such as $P_{human}$ and $P_{bf/bm \rightarrow \rm{criminal}}$) and model variants used.
It is evident that the \textit{brittleness} of these models certainly allows for trivially flipping the results to favor another narrative by smartly changing either the choice of labels, the choice of default-class (replacing \texttt{human being} with a synonym for example), the class sentence construction template or the model architecture variants (Using Vit-B/16/32 for example). Furthermore, parameters beyond our control (such as batch size used during pre-training, choice of tokenizer, and number of training epochs used) also likely played an important role in influencing these results.
Instead, what we are conveying through these results is simply this: despite making the most templated design choices on all the aspects of the pipeline, and despite verbatim replication of the empirical orchestration straight from the example code notebooks in the official Github repositories, and despite using an extremely controlled \textit{easy} probe dataset and class-design, \textit{it was verifiably hard to avoid the glaring negative impact on the biases measured that could be directly attributed to dataset scaling.}

Below we present a set of observations that we hope the ML community, dataset curators, as well as other stakeholders might find helpful towards advancing not only data curation but also the field as a whole in a manner that is transparent, rigorous, responsible, and accountable.  
\\\noindent\textbf{Avoid ad-hoc decision-making for dataset curation hyperparameters.} In the \textit{CLIP inference at the post-processing stage} section of the \href{https://laion.ai/blog/laion-5b/}{LAION-5B dataset announcement}, we encounter the fact that the dataset curators estimated the cosine similarity between an image and its alt-text description using the \texttt{ViT B/32} CLIP model and discarded all images with cosine similarity score of less than the manually set threshold of $0.28$. This is a marked departure from the procedure published during the \href{https://laion.ai/blog/laion-400-open-dataset/}{LAION-400M release} where the curators stated that \textit{``We use OpenAI’s CLIP model (the ‘ViT-B-32‘ version) to compute the image and alt text embeddings. Then we calculate the cosine similarity of both embedding vectors and drop all samples with a similarity below 0.3. We chose this threshold after trying different values and using human evaluations of how well the texts fit the images. Lower values like 0.28 or 0.29 also seemed okay in many cases, but after further inspections, we decided to choose the conservative value of 0.3''}. The reasoning behind this decision is not clear. However,  
such a decision might have been taken to boost the dataset size past the 5 billion mark, a pre-mandated milestone perhaps. Given these decisions have a significant consequence for dataset quality (subsequently model performance and potentially concrete lives through deployment), we recommend such processes be rigorously justified, well documented, and made transparent a la scientific practices.

\noindent\textbf{Beware of CFD physiognomy.}
Scholars have warned about the 
the rebirth of phrenology and physiognomy via the by-lanes of Computer Vision~\cite{stark2021physiognomic,spanton2022measuring}. 
Similarly, some of our preliminary investigations that emerged when we dug into the \textit{whyness} of criminality-association of some CFD faces by the models under consideration show  
high correlations with metrics such as Facial Width-to-Height Ratio (fWHR) and Cheekbone Prominence that are recorded as metadata in the CFD dataset. Well-informed and in-depth awareness of this pernicious development as well as mitigation mechanisms against phrenology is crucial. To this end, we encourage future research in line with those such as~\citet{hundt2022robots} 
to build upon this finding through a statistical experiment mapping the objective face-measurement-metrics found in `Study-1 and Table-1' of \cite{ma2015chicago} to the model outputs to further investigate  
the rebirth of phrenology and develop regulatory and mitigation mechanisms.
\\\textbf{Dataset sub-sampling: Only for ethics checks?}
There is an emergent trend within the broad culture of internal audits (self-audits within big corporations and institutes) focusing 
\textit{subsample-only-for-ethics-auditing} 
when it comes to handling large datasets, despite the abundant resources at their disposal. 
As far as training a monetizable model is concerned, scale is deemed a virtue and not a hindrance as exemplified by frequent aggressive crawl-scrape-scoop strategies. On the contrary,  
scale is deemed as  
an impediment when it comes to auditing, evaluating, and stress-testing datasets and models  
for critical concerns including checking for quality of data, encoded racial stereotypes, and bias. For example, we observed that the CLIP model was trained on a black-box Web-Image-Text (WIT) dataset spanning 400 million image text pairs. However, when it came to measuring the racial biases baked into the model, sub-sampling  was resorted to a 
comparatively \textit{small} dataset,  
the FairFace dataset~\cite{karkkainen2019fairface}, which only contains 
0.027\% (108,501 images) of the training dataset. 
Moreover, the bias-measurement exercise is minimal, limited only to 
running inference (read forward pass) through the model that is an order of magnitude less computationally intensive compared to training the model (backward pass). As stated in  
\textit{ Section 7.1: Bias} in the CLIP paper~\cite{clip_radford2021learning}, only 10000 images (0.0025\% of the training dataset size) were used from this FairFace dataset for the bias-check-inference task (that we have used in our experiments (see Section~\ref{sec:method})). We recommend audit, evaluation, and general critical and ethics work is carried out to the highest possible standards and scientific rigour. Otherwise, it risks ethics and audit washing. 
\\\textbf{Legal and policy implications.} The LAION datasets we audited serves as a critical backbone for numerous popular, influential and impactful models including Google's Imagen and Stable Diffusion variants. Increased integration of these models into numerous societal domains and practices means that these models are not purely intellectual exercises but result in direct or indirect impact on actual people, particularly marginalised groups. Yet, neither datasets nor information around dataset creation, curation, documentation, filtering and detoxifying mechanisms used are made available for most of these popular and influential models. Restricting access as well as active obfuscation of information around these datasets present a major obstacle to carrying out independent audits and developing appropriate regulatory guidelines and guardrails. Open access is a prerequisite to independent audits, particularly those aiming to examine, diagnose and challenge societal and historical injustices that datasets and AI models encode and amplify. We hope this work serves for legal and policy experts and authorities as a reminder for the urgent need to both encourage and develop legally enforceable mechanisms to allow access for independent audit and evaluation of training datasets. Our work also illustrates the importance of dataset curation, filtering and management. We highly recommend such practices become part and parcel of model development.

\section{Future Work and Conclusion}
\label{sec:conclusion}
We have carried out an extensive audit investigating the impact of dataset scale and model architecture on VLMs trained on the LAION datasets. 
In this regard, 
the emergence of projects such as \texttt{openclip}~\cite{ilharco_gabriel_2021_5143773_openclip} have been instrumental in  
allowing  for 
easy orchestration of the type of investigations executed and presented here. This section 
presents a list of natural extensions of our work. 
\\\textbf{BLIP and other CLIP models:}
In the associated GitHub repository, we have shared image-class cross-tabulated softmax matrices akin to the ones presented in Figure~\ref{fig:results:heatmap-images} for the other non-SoTA CLIP models presented in Table~\ref{tab:openclip_variants} for which we could run the \textit{fix-architecture-vary-training-datasets} experiments presented in the Results Section~\ref{sec:cfd_results}. We highly encourage  
for these experiments to be replicated across the other models including BLIP~\cite{li2022_blip} and the new variants emerging on the scene. We hope that this will help the ML community to intimately understand (and mitigate) the role that model architectures play in encoding harmful  
biases as the dataset scales.
\\\textbf{Choice of prompt template and class design:}
In this paper, we converted the categorical class labels into sentences using the format \textit{``A photo of $<$class$>$"} to maintain consistency with the CLIP~\cite{clip_radford2021learning} paper results. We posit that varying this prompt template with its rephrased variants such as \textit{``This a \textbf{picture} of $<$class$>$"} would  
result in variations of the results shown in Section~\ref{sec:cfd_results}. Similarly, we also expect that replacing the word \texttt{person} with the self-declared race-gender identifier (such as \texttt{asian-man}) will also result in variations to the cosine similarity value output by the models under consideration. Accordingly, future research might  
unearth the \textit{fairness-optimal} prompt template by both paraphrasings as well as choosing 
alternative-identifiers for the word \texttt{human being}.
\\\textbf{Extension across other expressions and other face datasets:}
In this paper, we have restricted our  
experimentation to the neutral expression images of the CFD dataset for the sake of brevity. One avenue for future work might be to investigate if holding the individuals' faces constant and varying the facial expressions makes a marked difference in the results. Also, inspired by the CFD project, we have seen the emergence of other similar datasets such as MR2~\cite{strohminger2016mr2}, Bogazici face database~\cite{saribay2018bogazici}, the Delaware dataset~\cite{mende2020delaware} and the ISIEA dataset~\cite{zheng2021isiea}. Replicating  
these experiments using these datasets might yield 
a more granular view of how these models -- supposedly trained on  
internet sourced data -- function and what 
biases might be baked into them. 
\\\textbf{The Race-Gender experiment: Some initial results:}
There also emerges the natural question with regards to the extent to which  stereotypes about facial appearances are cross-related with racial identities by these VLMs. Given that the CFD has self-classified race-gender labels, we also performed a small-scale race-gender classification experiment (similar to the FairFace experiment in the CLIP paper~\cite{clip_radford2021learning}), using the subjects' self-classified race-gender labels. That is, we replaced the 8 classes of [\texttt{human being},...,\texttt{suspicious person}] in the \textit{human-being experiment} above with the 8 self-classified race-gender category labels [\texttt{asian man},...,\texttt{white woman}]. The initial results are discussed in Appendix~\ref{app:rg_exp} and it appears as if faces with visible epicanthic folds (that occur across a broad spectrum of racial identities) are solely associated with the 'Asian' race identifier. This observation merits a deeper analysis especially given the wide availability of meta-data that is associated with the images in CFD that can be a rich source of confounding factors.

\subsection{Conclusion} 

We have carried out a dataset audit of two visio-linguistic multimodal datasets, LAION-400M and LAION 2B-en, and 14 Vision Transformer-based VLMs trained on them. We found evidence of misclassification in the models, particularly towards Black men and Latino men as `criminal', which exacerbates 
with training dataset size. We cannot stress the importance of open-source and in audit endeavors such as ours, since any kind of quantitative and qualitative dataset exploration hinges upon access to the artifacts themselves. We are saddened to see an increasing number of ML organizations fail to provide access to their datasets and models since we believe that this is an essential element to scientific advancement and a healthy, equitable, and innovative research community. 

Today’s state-of-the-art visio-linguistic multimodal models are trained with massive carbon footprints, massive data infrastructure, and massive funding. These models 
are currently being deployed in the real-world including in recommendation systems, information-retrieval systems, semantic search systems, and image captioning systems, although as we have illustrated in this paper, they can predict photographs of those with Black and Latin racial backgrounds as `criminal'.  
Given that such failures can result in dire consequences on real people, often those at the margins of society, we implore the research community as well as those developing and deploying these systems to carry out due diligence with rigorous audits of these models and training datasets and take necessary actions, including refraining from use in high-stake scenarios.

\begin{acks}
Abeba Birhane is supported by Science Foundation Ireland via the ADAPT Centre of Digital Content Technology funded under the European Regional Development Fund (ERDF) through Grant No \#13/RC/2106\_P2. Sepehr Dehdashtian and Vishnu Naresh Boddeti are partly supported by the National Science Foundation under Grant No. \#2147116. We would like to thank Ellen Rushe, Thomas Laurent, Andrew Hundt and the anonymous FAccT reviewers for their extensive and helpful feedback. 
\end{acks}

\section*{Ethical Considerations}
In this work, we have audited existing openly available datasets and VLM variants trained on them. We recognize these datasets pose numerous ethical concerns including the sourcing of these content that forms these datasets without consent, awareness or financial compensation for people in these datasets. The audit results are also disturbing and distressing, particularly to Black men who were predicted as ``criminal" with close to 100\% frequency for the model with patch size 32. We hope by bringing these to light, existing structures and systems of oppression can be challenged. In using the CFD as a probe dataset, we have been fully transparent about its limitations (see below) to help contextualise our findings. We believe our audit work poses no harm or risk to individuals or groups. To further minimise any potential privacy risk to individuals behind the CFD, we have hand-blurred all instances of the CFD throughout the paper.

\section*{Limitation}
The racial and gender construction and limited categories of the CFD adheres to gender and race essentialism. The binary gender (female or male) and the seemingly clean racial (Black, White, Asian, or Latin) categories used in the CFD fail to capture genders and races the real world presents. Far from these binary categories, genders and races are fluid, complex, multivalent, and multidimensional in actuality. Furthermore, as~\citet{hundt2022robots} point out, the individuals of CFD were provided with limited pre-defined categories (as opposed to given the agency to self-identify) to select their identities from. Yet, despite this limitation, we believe the dataset presents a useful proxy in the context of our experiments.

Additionally, we also details 
four sources of confounding factors that scholars investigating these biases need to consider that are beyond the scope of the work published here.
\begin{compactitem}
    \item Shortcomings of the cosine similarity metric during dataset curation process
    \item CLIP-like models suffering from Concept Association Bias (CAB)
    \item CLIP-like models exhibiting Bags-Of-Words like behavior
    \item CLIP-like models being vulnerable to Identity Inference Attack (IDIA) 
\end{compactitem}
\subsection{Effect of cosine similarity metric during dataset curation}
During the dataset curation stage of LAION datasets, cosine similarity between the text and image embeddings has been used to filter images that had \textit{reasonable} textual explanation associated with them in the alt-text field. It was hand-set to $0.3$ during the LAION 400M curation process and reduced to $0.28$ during the LAION-5B curation process. It has recently come to light by the work presented in~\citet{steck2024cosine} that this metric of  cosine similarity can potentially yield \textit{“arbitrary and therefore meaningless {similarities}”} for the learned embeddings in deep models such as CLIP and this constitutes another source of caution for future dataset curation practices.
\subsection{CLIP-like models suffering from Concept Association Bias (CAB)}
CLIP-like VLMs have been shown to suffer from Concept Association Bias (CAB)~\cite{tang2023lemons} on account of being trained on contrastive losses (in lieu of autoregressive losses).  Recently, \citet{tang2023lemons} have uncovered an interesting behavior that they termed as Concept Association Bias (CAB) that resulted in VLMs treating inputs as a bag of concepts and attempting to fill in the other missing concept crossmodally often resulting in unexpected zero-shot prediction. We belive that it will be an intriguing downstream study to critique our work and disentangle the contribution of this CAB that resulted in racially biased results.
\subsection{Bags-Of-Words like behavior}
CLIP-like VLMs fail to encode the compositional relationships between objects and attributes in the images thus  displaying Bags-Of-Words like behavior~\cite{yuksekgonul2022and}.
Similar to \citet{tang2023lemons}, 
~\citet{yuksekgonul2022and} also discovered the settings in which CLIP-like VLMs treated the constituent objects in an input image as bags-of-words thus displaying limited relational understanding and order insensitivity. They also advocated for composition-aware hard negative mining (CAHNM) as a potential solution  and juxtaposing the performance of plain-vanilla VLMs and CAHNM-improved VLMs  will be an interesting vector of research exploration.
\subsection{Data-leakage and Identity Inference Attacks (IDIA)}
CLIP-like large scale VLMs are presented with 
millions of human images during training and thus are vulnerable to data-leakage and Identity Inference Attacks (IDIA).
When we run benchmarking tests on VLM models that have been trained on internet-scale datasets, there exists a real possibility that the individuals who appear in the probe-test datasets might well have appeared in the VLM’s training dataset. In a recent paper titled \textit{“Does CLIP Know My Face?”},  ~\citet{hintersdorf2022does}  
empirically demonstrated how VLMs trained on the LAION-400M dataset trapped and leaked information about individuals appearing less than 25 times in the dataset and how one could preemptively check for this before finalizing on the probe images.
We have not done any data-leakage checks in this work and it is an interetsing topic to explore in a future dissemination.
\section*{Positionality Statement} 
We acknowledge any research process and subsequent knowledge produced cannot be entirely separable from various structural, institutional and personal factors. Seemingly invisible influencing factors include current trends in the field, interests of funding bodies, availability of resources, as well as the interests, motivations, goals, perspectives, and backgrounds of the researchers themselves. Thus, acknowledgement of these factors and transparency (and not hiding behind the veil of objectivity) is instrumental for research excellence. Our team is multi-racial and multi-gender and includes graduate and post-graduate researchers, senior researcher and an independent researcher engaged with AI, machine learning, computer vision, cognitive science, critical race theories, and algorithm and data audits. Having said that, we may have gaps in representing what might be most important to communities at the margins of society. Furthermore, we are all housed within Western universities, a privilege which enabled us to carry out and publish this research with relative ease compared to our peers who may not have the resources or compute available to carry out similar work, for example, those in non-Western universities. 

\bibliographystyle{ACM-Reference-Format}
\bibliography{facct24-83}

\clearpage 
\appendix
\section*{Appendix}

\section{Additional methodological details}

\subsection{Self-similarity matrix of CFD extracted featured}
\begin{figure*}[h!]
    \centering
    \includegraphics[width=0.8\columnwidth]{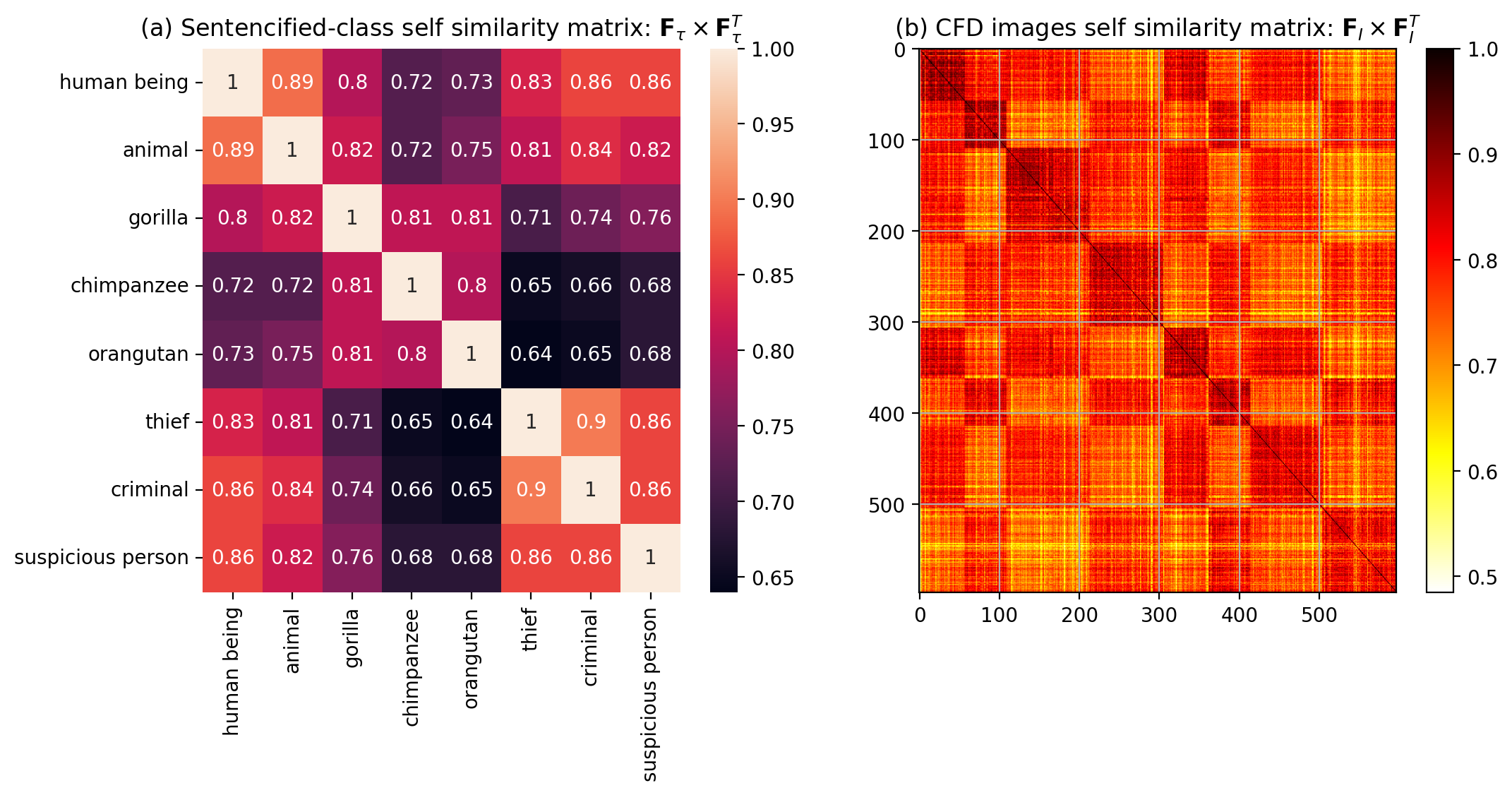}
    \caption{ Heatmap plots to help the reader visualize the (a) self-similarity matrix: $\mathbf{F}_\tau \times \mathbf{F}_\tau^T$ of the sentences corresponding to the class-labels and (b) self-similarity matrix: $\mathbf{F}_I \times \mathbf{F}_I^T$ of the features extracted from the CFD images.} 
    \label{fig:self_sim}
\end{figure*}

To highlight how self-similar the $8 \times 8$ textual features are, we present Figure~\ref{fig:self_sim}(a) that has the annotated heatmap of the $\mathbf{F}_\tau \times \mathbf{F}_\tau^T$ matrix. Similarly, we also present Figure~\ref{fig:self_sim}(b) that has the heatmap of the $597 \times 597$ sized $\mathbf{F}_I \times \mathbf{F}_I^T$ matrix. Given the fact that the 597 images were sorted and grouped by Race-Gender categories, the block-like structures visible in Figure~\ref{fig:self_sim}(b) indicate the fact that the model's output image features are 
influenced by these categorical indicators.

\subsection{Randomly selected, hand-blurred samples from the CFD}
\begin{figure*}[!ht]
    \centering
    \includegraphics[width=0.8\columnwidth]{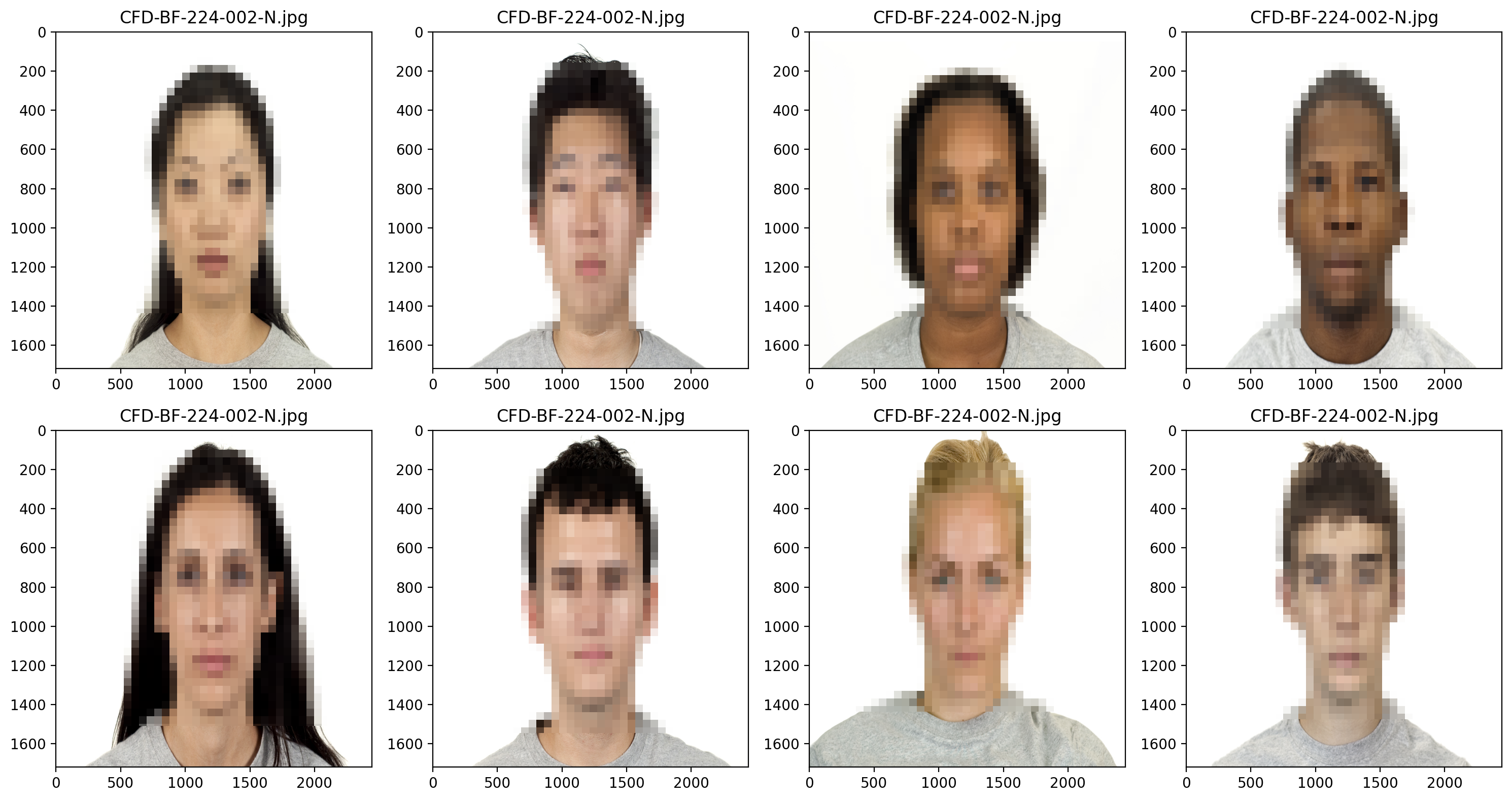}
\caption{
A sample of images from the Chicago Face Database (CFD) across the eight self-classified race-gender combinations.}
\label{fig:cfd-images}
\end{figure*}
A sample of images from the Chicago Face Database (CFD) across the eight self-classified race-gender combinations. The images are sized $2444(w) \times 1718(h)$ pixels and \textit{``equated for color temperature and placed onto a plain white background"}. Of the 597 individuals,  307 self-classified as ``female" and 290 self-classified as ``male". We hand-blurred these sample images for this study to preserve the anonymity of pictured individuals. The titles of each of these images here follow the exact file names given to these images in the \texttt{CFD 3.0} version that is hosted at \url{https://www.chicagofaces.org/download/}.

\section{On AllLookSameism, negative stereotypes and racial misclassification}
\label{app:rg_exp}
\begin{figure*}[!ht]
    \centering
    \includegraphics[width=\textwidth]{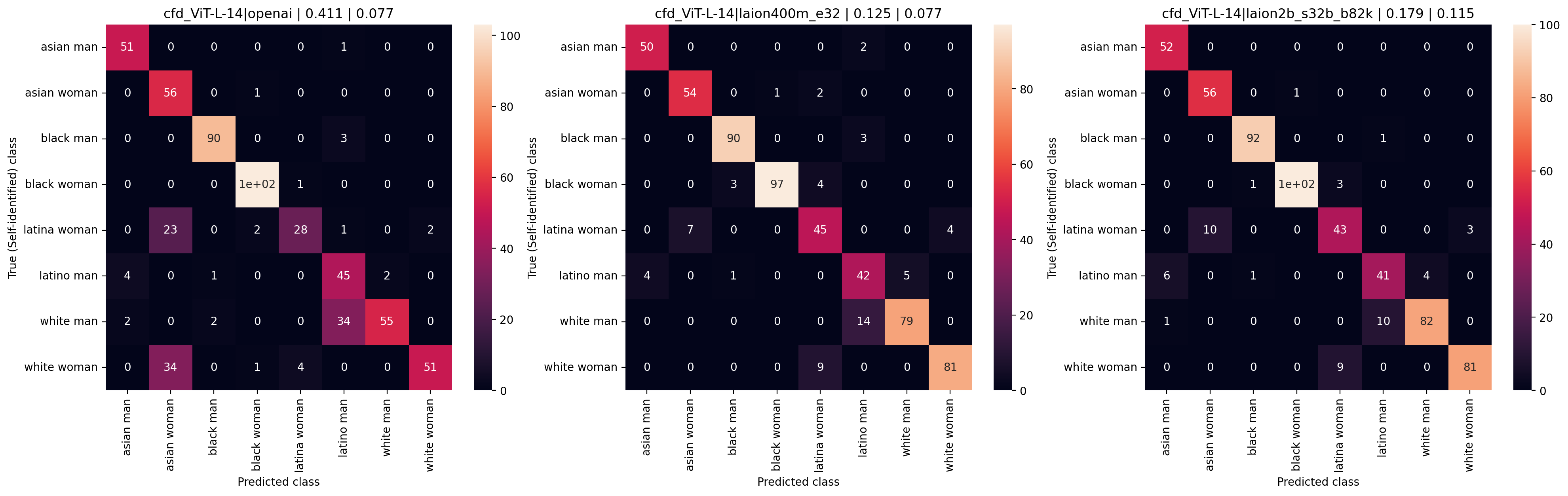}
    \caption{Heatmap of the confusion matrix of the race-gender classification experiment showing misclassification Latino/Latina individuals as `Asian' class. This misclassification got worse with dataset scaling.}
    \label{fig:crosstab_rg}
\end{figure*}
The goal here was to understand how stereotypes about facial appearances are cross-related with racial identities. When we looked at the results (Figure~\ref{fig:crosstab_rg}) we saw an interesting theme emerge: the self-classified Latino/Latina individuals were misclassified with high confidence as one of the `Asian' classes on account of the presence of \textit{epicanthic folds} and this tendency to stereotype got worse with dataset scaling.
The titles of these subplots here are formatted as strings with 4 fields separated by the `$|$' character:
$<\text{cfd\_Vit-L-14}>|<\text{training-dataset}>|<P_{lf \rightarrow af}>|<P_{lm \rightarrow am}>$. Here, $P_{lf \rightarrow af}$ is the probability that an image belongs to the \texttt{Latina-Female} category was misclassified as \texttt{Asian-Female} ( and $P_{lm \rightarrow am}$ is the probability that an image belongs to the \texttt{Latino-Male} category was misclassified as \texttt{Asian-Male}). As seen in the first of the 3 subplots (from left) that maps to the OpenAI-WIT dataset 23 of the 56 Latina women were misclassified as Asian women leading to a $P_{lf \rightarrow af}=23/56=0.411$. This misclassification rate was better for the LAION-400M model ($0.125$) and worsened to $0.179$ for the LAION-2B-En model, thereby yielding yet another example of worsening of the bias-related metrics upon scaling the dataset from 400M to 2B samples. The same trend also showed up for Latino men with the misclassification rate increasing nearly $50\%$ from $0.077$ to $0.115$.
\\
Correspondingly, there exists a substantial body of scientific literature (See \cite{fry2017latinx,gross2009own_latinx, pacheco2008rhetoric_latinx,nakamura2005alllooksame_latinx}) on not just the oft-ignored high levels of prevalence of the epicanthic folds in Hispanic/LatinX populations\footnote{``In Latinos, the inner canthal distance and lateral canthal angle of inclination were similar to Asians, while the lid crease spanned the range from Asians to Caucasians. Half of the Latinos had epicanthal folds"~\cite{fry2017latinx}} but also on the sociological ramifications of this \textit{alllooksame-ism}~\cite{nakamura2005alllooksame_latinx} that permeates aspects of the mainstream culture.

\end{document}